\documentclass[%
reprint,
superscriptaddress,
amsmath,amssymb,
aps,
]{revtex4-2}
\usepackage{amsmath}
\usepackage{amssymb}
\usepackage{graphicx}
\usepackage{braket}
\usepackage{color}
\usepackage{siunitx}

\usepackage[utf8]{inputenc}
\usepackage{physics}
\usepackage{multirow}
\usepackage{url}
\usepackage{dcolumn}% Align table columns on decimal point
\usepackage{bm}% bold math
\begin{document}

%\makeatletter
% \renewcommand\@biblabel[1]{#1.}
%\makeatother

%%%%%%%%%%%%%%%%%%%%%%%%%%%%%%%%%%%%%%%%
%%%%%%% TITLE & AUTHORS
%%%%%%%%%%%%%%%%%%%%%%%%%%%%%%%%%%%%%%%%

\title{Spin-relaxation times exceeding seconds for color centers with strong spin-orbit coupling in SiC}

\author{Carmem~M.~Gilardoni}\thanks{These authors contributed equally to this work.}
\author{Tom~Bosma}\thanks{These authors contributed equally to this work.}
\author{Danny~van~Hien}
\author{Freddie~Hendriks}
\affiliation{Zernike Institute for Advanced Materials, University of Groningen, NL-9747AG  Groningen, The Netherlands}
\author{Bj\"orn~Magnusson}
\affiliation{Norstel AB, Ramsh\"allsv\"agen 15, SE-602 38 Norrk\"oping, Sweden}
\author{Alexandre~Ellison}
\affiliation{Norstel AB, Ramsh\"allsv\"agen 15, SE-602 38 Norrk\"oping, Sweden}
\author{Ivan~G.~Ivanov}
\affiliation{Department of Physics, Chemistry and Biology, Link\"oping University, SE-581 83 Link\"oping, Sweden}
\author{N.~T.~Son}
\affiliation{Department of Physics, Chemistry and Biology, Link\"oping University, SE-581 83 Link\"oping, Sweden}
\author{Caspar~H.~van~der~Wal}
\affiliation{Zernike Institute for Advanced Materials, University of Groningen, NL-9747AG  Groningen, The Netherlands}

\date{Version of \today}

% Correspondence: C. M. Gilardoni (c.maia.gilardoni@rug.nl) or C.H. van der Wal (c.h.van.der.wal@rug.nl)

\begin{abstract}

Spin-active color centers in solids show good performance for quantum technologies. Several transition-metal defects in SiC offer compatibility with telecom and semiconductor industries. However, whether their strong spin-orbit coupling degrades their spin lifetimes is not clear.  We show that a combination of a crystal-field  with axial symmetry and spin-orbit coupling leads to a suppression of spin-lattice and spin-spin interactions, resulting in remarkably slow spin relaxation. Our optical measurements on an ensemble of Mo impurities in SiC show a spin lifetime $T_1$ of $2.4~\text{s}$ at $2~\text{K}$.

\end{abstract}

\maketitle

%%%%%%%%%%%%%%%%%%%%%%%%%%%%%%%%%%%%%%%%
%%%%%%% INTRODUCTION
%%%%%%%%%%%%%%%%%%%%%%%%%%%%%%%%%%%%%%%%
Spin-active color centers in semiconductors have attracted significant interest for the implementation of quantum technologies, since several of these systems combine long-lived spin states with a bright optical interface \cite{weber2010,Awschalom2013,aharonovich2016,atature2018}. Long distance spin entanglement has been achieved for a variety of defects as stationary nodes \cite{gao2015,hensen2015,Sipahigil2014,klimov2015}. However, finding suitable emitters that combine long-lived spins, short excited-state lifetimes and optical transitions compatible with telecommunication fiber-optics infrastructure in an industrially established material has remained elusive. Silicon carbide, a wide band-gap semiconductor with mature fabrication technology, hosts a range of defect centers with optical transitions near or at the telecom range \cite{magnusson2005,zargaleh2016}, including several defects containing transition metal (TM) impurities  \cite{gallstrom2015,koehl2017,bosma2018,spindlberger2019,csore2019,wolfowicz2019,diler2019}. The electronic and spin properties of these defects derive largely from the character of the d-orbitals of the TM under the action of a crystal field determined by the lattice site \cite{baur1997,kaufmann1997,abragam1970}. Furthermore, the presence of a heavy atom in the defect implies that spin-orbit coupling (SOC) plays a significant role in the electronic structure of these color centers. Generally, the presence of strong SOC is expected to degrade spin properties by coupling spin and orbital degrees of freedom \cite{diler2019}. However, the extent to which the spin lifetimes of TM defects are limited by the strong influence of SOC is not clear. 

We report here on slow spin relaxation with $T_1$ exceeding seconds below $4~\text{K}$ for a Mo defect ensemble in $6$H-SiC, indicating that the defect spin is surprisingly robust with respect to spin relaxation despite the presence of strong SOC. In order to understand this, we measure the spin-relaxation time of the Mo defect in SiC between $2$ and $7$ K and identify the main processes leading to spin relaxation in this temperature range. We analyze the manifestation strength of these processes, while considering the electronic structure of the defect, and find that SOC is in fact responsible for suppressing several spin-relaxation mechanisms in this system, leading to unexpectedly long $T_1$.

%%%%%%%%%%%%%%%%%%%%%%%%%%%%%%%%%%%%%%%%
%%%%%%% I) MATERIALS and METHODS
%%%%%%%%%%%%%%%%%%%%%%%%%%%%%%%%%%%%%%%%

\begin{figure}[h!]
	%\centering
	\includegraphics[width=8cm]{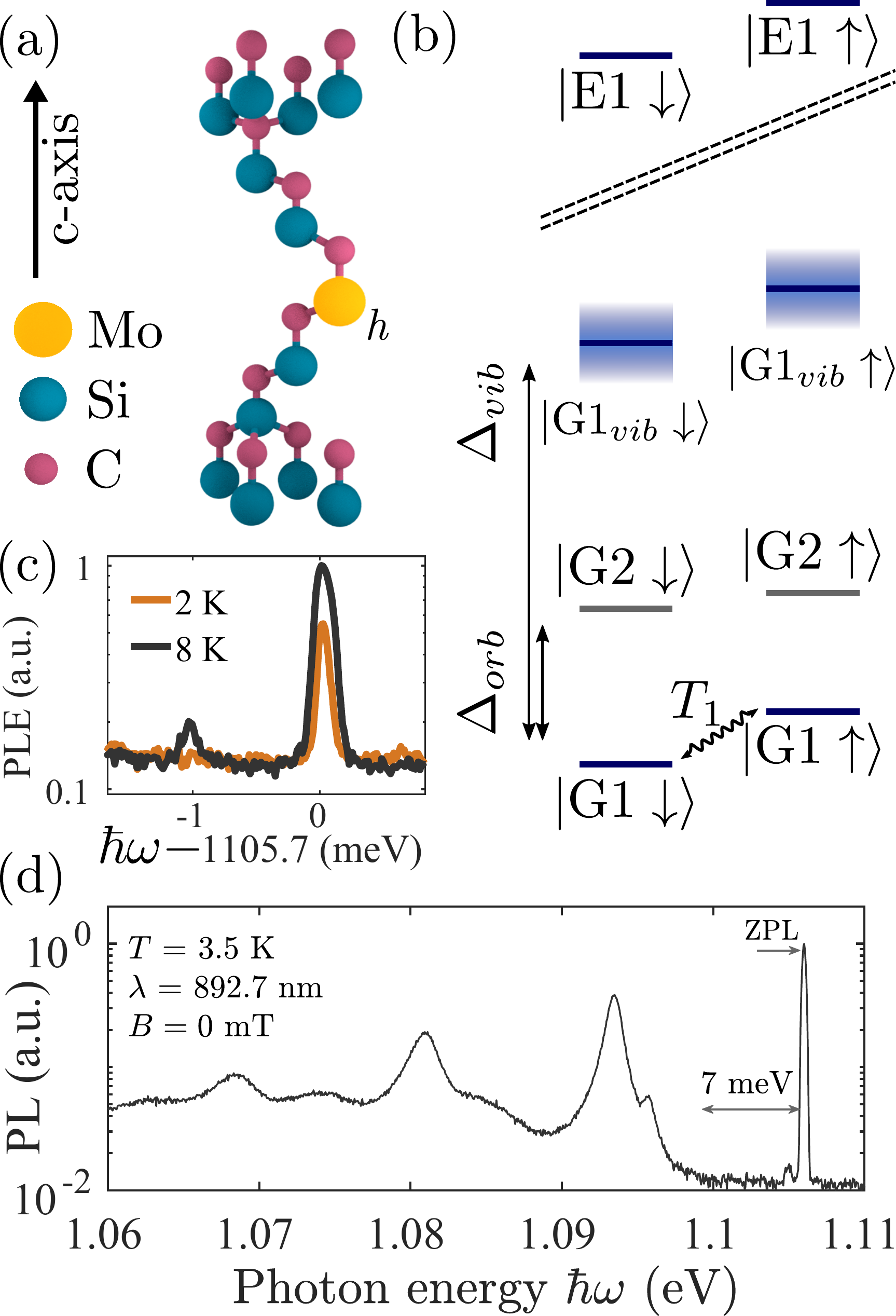}\\
	\caption{\textbf{Defect structure and electronic structure.} (a) The defect is composed of a Mo atom positively ionized substituting a Si at a quasi-hexagonal ($h$) lattice site. (b) Both ground ($\ket{\text{G}1}$) and optically excited ($\ket{\text{E}1}$) states are characterized by a spin $1/2$, and are split by a magnetic field. Additionally, a second ground-state Kramer's doublet ($\ket{\text{G}2}$) is present approximately $1~\text{meV}$ above the ground state doublet. (c) When this second doublet is thermally occupied, a second zero-phonon line (ZPL) appears in the photoluminescence excitation (PLE) spectrum, where the light emitted in the phonon sideband is measured as a function of the laser photon energy $\hbar \omega$. (d) The ground electronic state can couple to localized vibrational modes, giving rise to vibronic states $\ket{\text{G}1_{vib}}$, as evidenced by the phonon replicas observed in the photoluminescence (PL) spectrum.}
	\label{fig:Fig1}
\end{figure}

We focus on the Mo defect associated with an optical transition line at $1121.3~\text{nm}$ in $6$H-SiC, which consists of a Mo impurity in a Si substitutional site of quasi-hexagonal symmetry (Fig.~\ref{fig:Fig1}(a), $h$ site) \cite{bosma2018, csore2019}. The defect is positively ionized, such that after binding to four neighboring carbons, the TM is left with one active unpaired electron in its $4$-d shell. In this lattice site and only considering the rotational symmetry of the defect, both ground and excited states are two-fold orbitally degenerate. Due to this degeneracy, the orbital angular momentum is not quenched \cite{abragam1970}. In the presence of spin-orbit coupling, the orbital degeneracy is broken, giving rise to two Kramer's doublets (KD) in the ground and two KD's in the excited state (Fig.~\ref{fig:Fig1}(b)) \cite{bosma2018,csore2019}, resembling what is observed for the silicon-vacancy in diamond \cite{hepp2014}. Concerning its electronic structure, this defect shares several similarities with the V defect in SiC, which has optical transitions fully compatible with telecommunications infrastructure \cite{baur1997,bosma2018,wolfowicz2019}

Each KD is a doublet composed of a time-reversal pair: the doublet splits as an effective spin-$1/2$ system in the presence of a magnetic field, but its degeneracy is otherwise protected by time-reversal symmetry (see \cite{SI} for a more extensive summary). The energy difference between the two spin-orbit split KD's in the ground state, $\Delta_{orb}$ in Fig.~\ref{fig:Fig1}(b), is expected to be approximately $1$ meV \cite{csore2019}. This is in accordance with the appearance of a second zero-phonon line (ZPL) in resonant photoluminescence excitation (PLE) experiments at approximately $8~\text{K}$ (Fig.~\ref{fig:Fig1}(c)). Finally, the presence of sharp phonon replicas of the ZPL in the photoluminescence spectrum indicates that the defect center couples strongly to localized vibrational modes (Fig.~\ref{fig:Fig1}(b,d)).
 
\begin{figure}[h!]
	%\centering
	\includegraphics[width=8cm]{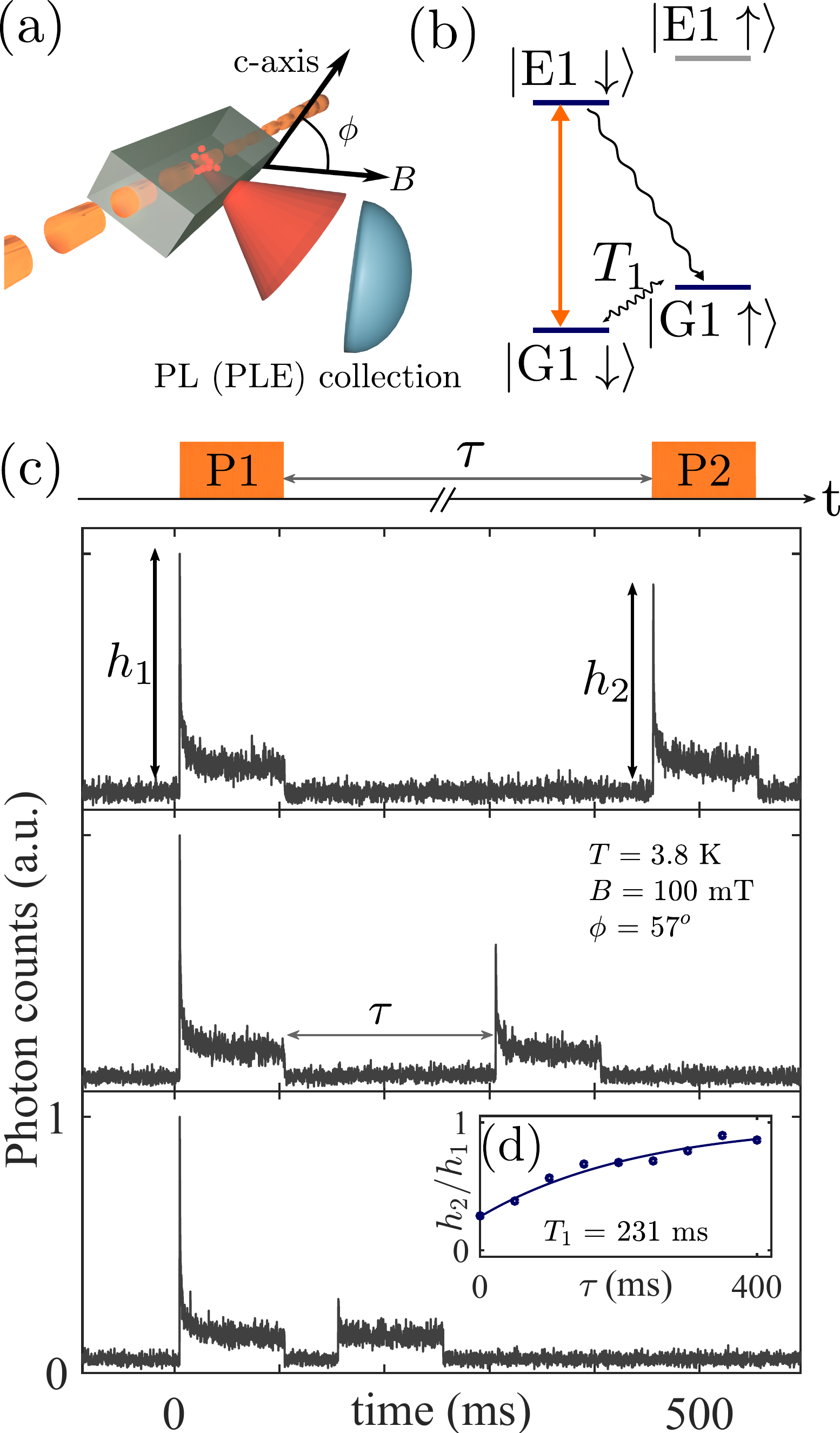}\\
	\caption{\textbf{Time-resolved hole burning experiment.} (a) In order to investigate $T_1$, we perform time-resolved pump-probe experiments where we resonantly excite the sample in the presence of a magnetic field and collect the phonon sideband emission. (b) When the magnetic field is non-collinear with the symmetry axis of the defect, spin-flipping transitions between ground and orbitally excited states are allowed. (c) The first pulse P$1$ probes the population at thermal equilibrium in the ground state spin sublevel $\ket{\text{G}1\downarrow}$ (which we call bright state in this work), subject to resonant excitation into the optically excited state. This pulse polarizes the spin population into the spin sublevel which is not subject to resonant optical excitations (which we here call dark state). After a variable delay $\tau$, a second pulse P$2$ probes the recovery of the population of the bright state. (inset) The ratio of the leading edge peaks in the PLE signal ($h_2$/$h_1$) recovers monoexponentially as a function of the delay between the two pulses, such that we can extract $T_1$.}
	\label{fig:Fig2}
\end{figure}

We measure the spin relaxation time of this defect by means of a pump-probe experiment as shown in Fig.~\ref{fig:Fig2}. Experiments are performed on an esemble of defects in a $6$-H SiC sample previously investigated in Ref. \cite{bosma2018}, and further described in the SI \cite{SI}. We create pulses out of a CW laser beam by using a combination of an electro-optical phase modulator (EOM) and a Fabry-Pérot (FP) cavity. The EOM generates sidebands from our CW laser at frequency steps determined by an RF input signal; by tuning the FP cavity to transmit this sideband only, we create pulses that turn on/off as the RF generator turns on/off. These pulses resonantly drive optical transitions between ground and excited states, and we measure the photon emission into the phonon sideband with a single-photon counter after filtering out the laser line. We apply a magnetic field non-collinear with the c-axis of the sample such that spin-flipping transitions between ground and optically excited states are allowed (Fig.~\ref{fig:Fig2}(a-b)) \cite{bosma2018}. In order to counteract slow ionization of the defects (see \cite[Sec. III]{SI} for further details), we apply a repump laser in between measurements \cite{beha2012,wolfowicz2017,zwier2015,bosma2018}. 

%%%%%%%%%%%%%%%%%%%%%%%%%%%%%%%%%%%%%%%%%%%
%%%%%%% II) T1
%%%%%%%%%%%%%%%%%%%%%%%%%%%%%%%%%%%%%%%%%%%

In the pump-probe experiment (Fig.~\ref{fig:Fig2}(c)), the initial response of the sample to pulse P$1$ provides a measure of the population in the bright spin state $\ket{\text{G}1\downarrow}$ at thermal equilibrium (see the caption of Fig.~\ref{fig:Fig2} for an explanation of the terms dark and bright spin sublevels in this work). The sharp increase (decrease) of the PL signal as the pulse turns on (off) indicates that both the optical decay rate and the Rabi frequency are relatively fast (see \cite{SI} for further details). In fact, an orbital decay rate of the order of a few MHz, in agreement with previous estimates \cite{bosma2018}, would explain the absence of a PL tail as the laser pulse turns off. Optical excitation provided by P$1$ polarizes the spin ensemble in a dark spin sublevel ($\ket{\text{G}1\uparrow}$ in Fig.~\ref{fig:Fig2}(b)) within the first few microseconds, as evidenced by the decrease and subsequent saturation of the PLE signal. The leading-edge response of the ensemble to a second (probe) pulse P$2$ reflects the recovery of the population in the bright spin sublevel ($\ket{\text{G}1\downarrow}$ in Fig.~\ref{fig:Fig2}(b)) during the time $\tau$ between the two optical pulses. Between $2~\text{K}$ and $7~\text{K}$, we observe a monoexponential recovery of $h_2$ towards $h_1$ as a function of $\tau$ (Fig.~\ref{fig:Fig2}(d)), which must correspond to the spin relaxation time $T_1$ given the considerations presented below.

We repeat this experiment at zero magnetic field in order to confirm that the PL darkening observed within the first few $\mu$s of optical excitation corresponds to optical pumping of the ensemble into a dark spin sublevel. In this case, we observe no leading-edge peak in the photoluminescence, indicating that no optical pumping occurs in this timescale if the spin sublevels are degenerate \cite[Fig. S$2$]{SI}. The absence of PL darkening at zero magnetic field implies that we cannot trap Mo centers in state $\ket{\text{G}2}$ for observable times. This is the case if the optical decay rate between states $\ket{\text{E}1}$ and $\ket{\text{G}2}$ is smaller than the relaxation rate between the two ground state KD's $\ket{\text{G}1}$ and $\ket{\text{G}2}$ in the temperature range investigated. (We do expect the optical decay rate between states $\ket{\text{E}1}$ and $\ket{\text{G}2}$ to be small, since we observe no lines corresponding to this transition in PL and PLE scans.) Thus, we conclude that after optical pumping in the presence of a magnetic field, no significant population is trapped in state $\ket{\text{G}2}$, and that relaxation dynamics via this level does not affect the signals used to derive $T_1$. Furthermore, we investigate the timescale associated with bleaching due to ionization \cite{wolfowicz2017,zwier2015}, which is found to be several orders of magnitude slower than the one associated with spin dynamics \cite[Sec. III]{SI}.

%%%%%%%%%%%%%%%%%%%%%%%%%%%%%%%%%%%%%%%%%%%
%%%%%%% II.a) Temperature Dependence
%%%%%%%%%%%%%%%%%%%%%%%%%%%%%%%%%%%%%%%%%%%

\begin{figure}[h!]
	%\centering
	\includegraphics[width=8cm]{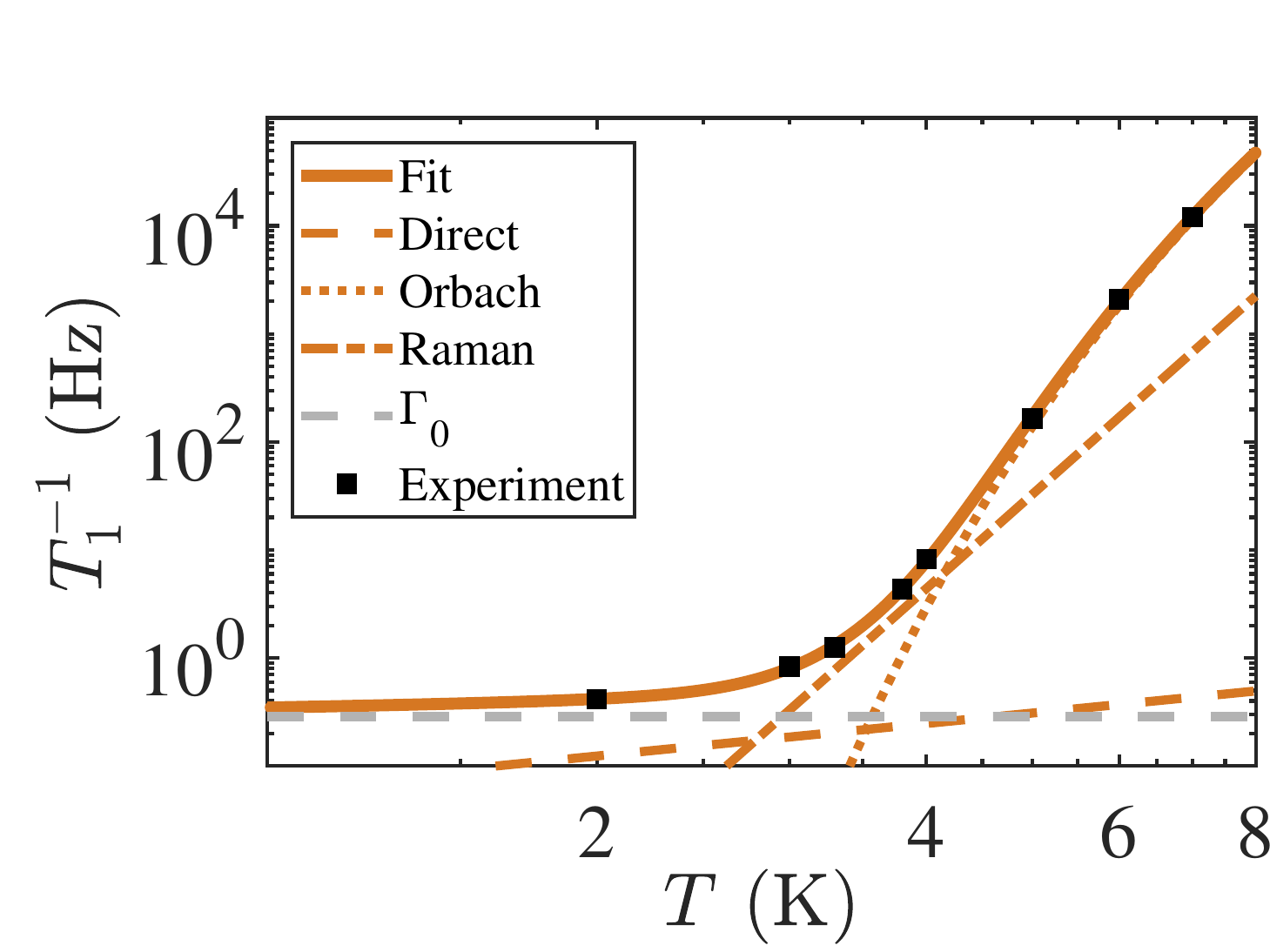}\\
	\caption{\textbf{Temperature dependence of spin relaxation.} The variation of $T_1^{-1}$ with temperature can be accurately described by a combination of direct (one phonon), Raman and Orbach processes (two-phonon processes). The two-phonon processes happen due to coupling to an excited state approximately $7$ meV above the ground state KD, compatible with coupling of the electronic state of the defect to localized vibrational modes (see main text).}
	\label{fig:Fig3}
\end{figure}

The temperature dependence of the spin relaxation time spans several orders of magnitude, going from $2.4~\text{s}$ at $2$ K to $83$ $\mu$s at $7$ K (Fig.~\ref{fig:Fig3}). Concerning phonon-mediated spin relaxation, the mechanisms leading to spin relaxation are well established \cite{abragam1970,stevens1967}. One (direct) and two-phonon (Raman, Orbach) processes are relevant when transitions between the levels involved are thermally accessible. Direct spin flip processes are expected to lead to spin relaxation rates that grow linearly with temperature ($T_{1,direct}^{-1} \propto T$). In contrast, two phonon processes give rise to spin relaxation rates that grow superlinearly with temperature ($T_{1,Raman}^{-1} \propto T^{5<n<11}$ and $T_{1,Orbach}^{-1} \propto e^{-\Delta/k_B T}$, where $\Delta$ is the energy of the relevant excited state) \cite{abragam1970, stevens1967, kiel1967, shrivastava1983}. Additionally, interaction with paramagnetic moments in the material are expected to lead to temperature independent spin-relaxation. We can fit the data presented in Fig.~\ref{fig:Fig3} to a combination of these processes and identify the temperature regimes where they are relevant (see \cite[Sec. VII]{SI} for additional details concerning the fitting procedure). 

The defect has a rich electronic structure, with orbital and vibrational degrees of freedom at energies that are thermally accessible between $2$ and $7$~K (Fig.~\ref{fig:Fig1}(b)). Thus, we expect two-phonon processes involving real (Orbach) and virtual (Raman) transitions into both vibrational and orbital excited states to contribute to the temperature dependence of $T_1^{-1}$. Surprisingly, however, we find that not all available states contribute significantly to spin relaxation, leading to unexpectedly long spin lifetimes below $4~\text{K}$. 

Above $4$~K, we identify an exponential growth of $T_1^{-1}$ as a function of temperature, indicating the prevalence of spin-relaxation via Orbach processes in this temperature range. From the fit presented in Fig.~\ref{fig:Fig3}, we extract $\Delta = 7 \pm 1~\text{meV}$ for the energy of the excited state involved in these spin flips. This energy matches the difference observed in PL experiments between the zero-phonon line and the onset of the phonon-sideband emission (Fig.~\ref{fig:Fig1}(d)). From the PL spectrum, we extract $\Delta_{vib} \approx 10~\text{meV}$ for the energy of the first phonon replica (Fig.~\ref{fig:Fig1}(c)); nonetheless, the phonon replicas are significantly broadened, such that the first available vibrational levels are lower in energy. Thus, we identify vibronic levels where the spin is coupled to localized vibrational modes of the defect as the relevant excited state for two-phonon mediated spin relaxation processes. As expected, these spin-relaxation mechanisms do not depend on the magnitude of the magnetic field \cite{SI}. 

We note that we cannot observe Orbach processes involving states $\ket{\text{G}2}$. Between $2$ and $7$~K, the contribution of these processes to the spin relaxation is expected to be close to saturation, giving a flat temperature dependence in this range. Nonetheless, the long spin relaxation times observed at $2$~K provide an upper bound to the spin relaxation rates due to two-phonon processes mediated by $\ket{\text{G}2}$, and are evidence that these processes are slow. This leads us to conclude that transitions between these two KD's are strongly spin-conserving, similarly to what is observed for SiV color centers in diamond \cite{rogers2014,becker2018}. The presence of the additional orbital degree of freedom does not significantly degrade the spin relaxation time of this defect since hopping between states $\ket{\text{G}1}$ and $\ket{\text{G}2}$ preserves the spin state (but does contribute to decoherence \cite{jahnke2015}). 

Below $4$~K, the slow spin-relaxation rates observed must result from three mechanisms: \emph{i}) processes whose temperature dependence is saturated in this range (as treated above); \emph{ii}) direct one phonon transitions between the two spin sublevels $\ket{\text{G}1\downarrow}$ and $\ket{\text{G}1\uparrow}$ and \emph{iii}) temperature independent processes (such as spin-spin paramagnetic coupling). The observation of $T_1$ above seconds below $4~\text{K}$ is evidence that all three processes must be relatively slow for this particular system. The first contribution has been discussed above. Below, we elaborate on how the electronic structure of this defect leads to a suppression of the latter two processes. 

The two spin-sublevels pertaining to a KD are strictly time-reversal symmetric with respect to each other, such that they must be degenerate eigenstates of any external field that preserves time-reversal symmetry. For this reason, electrostatic fields arising from the presence of phonons cannot, to first order, cause a direct spin-flip in a one-phonon type of transition between pure KD pairs \cite[Sec. IV]{SI,abragam1970}. A magnetic field or the interaction with nearby nuclear spins breaks time-reversal symmetry, and leads to mixing between spin sublevels pertaining to different KD. It is only due to this mixing that direct spin-flips via interactions with single phonons can occur. This mixing is inversely proportional to the energy separation between the various KD, which is in turn largely determined by the spin-orbit splitting. Thus, a large spin-orbit coupling protects the KD character of the ground state spin doublet, suppressing direct spin-flipping processes.

Additionally spin-orbit coupling leads to a highly anisotropic Zeeman splitting of the ground-state spin sublevels \cite{bosma2018,SI}, which hinders its interaction with a bath of paramagnetic impurities in the SiC crystal. Firstly, the spins are insensitive to magnetic fields perpendicular to their quantization axis, such that small fluctuations in the local magnetic field are not likely to induce a spin flip. Secondly, the spin doublet always has a Zeeman splitting governed by a $g$ factor that is at maximum $1.6$ \cite{bosma2018}, such that resonant spin flip-flop interactions with neighboring paramagnetic impurities with $g \approx 2$ are largely suppressed. 

The same arguments presented above to explain the long spin-relaxation times observed in this defect indicate that obtaining control of the defect spin via electric or magnetic microwave fields is a challenge. In first order, the ground state electronic spin is not allowed to couple to magnetic fields perpendicular to its quantization axis, or to electric fields (see \cite[Sec-III]{SI} for a detailed group-theoretical analysis supporting this). As a consequence, one cannot directly drive spin resonances between the two ground state spin sublevels. Mixing of the ground state KD with states lying at higher energies due to the presence of magnetic, hyperfine or electric fields can, however, lift these restrictions as is observed for V defects in SiC \cite{wolfowicz2019}. Although this is expected to provide a contribution to spin relaxation mechanisms, it is also a necessary requirement in order to control the ground state electronic spin via microwaves.

%%%%%%%%%%%%%%%%%%%%%%%%%%%%%%%%%%%%%%%%%%%
%%%%%%% III) Discussion
%%%%%%%%%%%%%%%%%%%%%%%%%%%%%%%%%%%%%%%%%%%

As discussed above, in the particular case of a TM at a quasihexagonal site, such as the Mo center, a combination of rotational symmetry and strong SOC protects the effective spin from flipping via two independent mechanisms. On the one hand, it gives rise to a unique Zeeman structure that is insensitive to perpendicular magnetic fields and non-commensurate with the behavior of other paramagnetic centers in the crystal. On the other hand, SOC generates a  ground state effective spin which is an isolated KD, protected by time-reversal symmetry. In this doublet, the crystal field locks the orbital angular momentum along the axis of rotational symmetry of the defect. Via the strong spin orbit coupling, the electronic spin is stabilized, giving rise to robust effective spin states with long spin-relaxation times.

Considering these processes alone, stronger SOC is thus expected to lead to slower spin relaxation rates in TM color centers with an odd number of electrons and rotational symmetry. Nonetheless, our work shows that the presence of localized vibrational modes is pivotal in generating spin-flips, their energy determining the temperature where the onset of two-phonon relaxation mechanisms happens. Since the energy of the localized vibrational modes depends non-trivially on the reduced mass of the defect, whether or not defects containing heavier TMs (where SOC is more prevalent) will exhibit longer spin-relaxation times remains a question. Thus, it would be interesting to investigate these processes in defects containing $5$-d electrons in SiC, where SOC is enhanced further. Tungsten defects, for example, have been observed with an optical transition at approximately $1240~\text{nm}$ and odd number of electrons, although their particular lattice site and charge state are not fully determined yet \cite{gallstrom2012}.

%%%%%%%%%%%%%%%%%%%%%%%%%%%%%%%%%%%%%%%%%%%
%%%%%%% IV) Conclusion
%%%%%%%%%%%%%%%%%%%%%%%%%%%%%%%%%%%%%%%%%%%

More generally, SOC should not be regarded as a detrimental feature when investigating solid-state defects for quantum communication applications. Transition metal defects in SiC offer interesting opportunities, such as charge state switching \cite{kunzer1993, dalibor1997}, emission in the near-infrared \cite{spindlberger2019} and long spin-lifetimes \cite{diler2019}. The maturity of the SiC-semiconductor industry means that a wide range of defects has been identified \cite{baur1997}. Nonetheless, their characterization with respect to optical and spin properties is still vastly unexplored. 

%%%%%%%%%%%%%%%%%%%%%%%%%%%%%%%%%%%%%%%%
%%%%%%% Figures
%%%%%%%%%%%%%%%%%%%%%%%%%%%%%%%%%%%%%%%%

%%%%%%%%%%%%%%%%%%%%%%%%%%%%%%%%%
%%%%%%%%%% METHODS
%%%%%%%%%%%%%%%%%%%%%%%%%%%%%%%%%

%%%%%%%%%%%%%%%%%%%%%%%%%%%%%%%%%
%%%%%%%%%% DATA AVAILABILITY
%%%%%%%%%%%%%%%%%%%%%%%%%%%%%%%%%

%\vspace{1cm}
%\noindent \textbf{Data availability}\\
%\noindent The data sets generated and analyzed during the current study are available from the corresponding author upon reasonable request.

%%%%%%%%%%%%%%%%%%%%%%%%%%%%%%%%%
%%%%%%%%%% ACKNOWLEDGEMENTS
%%%%%%%%%%%%%%%%%%%%%%%%%%%%%%%%%
\clearpage

\vspace{1cm}
\noindent \textbf{Acknowledgements}\\ We thank A. Gali, A. Cs\'or\'e and P. Wolff for discussions. Technical support from H. de Vries, J. G. Holstein, T. J. Schouten, and H. Adema is highly appreciated. Financial support was provided by the Zernike Institute BIS program, the EU H2020 project QuanTELCO (862721), the Swedish Research Council grants VR 2016-04068 and VR 2016-05362, the Knut and Alice Wallenberg Foundation (KAW 2018.0071),  and the Carl Tryggers Stiftelse för Vetenskaplig Forskning grant CTS 15:339.

%%%%%%%%%%%%%%%%%%%%%%%%%%%%%%%%%%%%%%%%%%%%
%%%%%%%%%% COMPETING INTEREST STATEMENT
%%%%%%%%%%%%%%%%%%%%%%%%%%%%%%%%%%%%%%%%%%%%

%\vspace{1cm}
%\noindent \textbf{Competing interest}\\
%\noindent The authors declare no competing financial or non-financial interests.

%%%%%%%%%%%%%%%%%%%%%%%%%%%%%%%%%%%%%%%%%%%%
%%%%%%%%%% AUTHOR CONTRIBUTIONS
%%%%%%%%%%%%%%%%%%%%%%%%%%%%%%%%%%%%%%%%%%%%

\vspace{1cm}
\noindent \textbf{Author Contributions}\\ 
The project was initiated by C.H.v.d.W. and T.B. SiC materials were grown and prepared by A.E. and B.M. Experiments were performed by T.B., D.v.H. and C.G, except for the PL measurements which were done by  I.G.I. Data analysis was performed by C.G., T.B., D.v.H., F.H. and C.H.W. C.G., T.B., and C.H.W. had the lead on writing the paper, and C.G. and T.B. are co-first author. All authors read and commented on the manuscript.

\clearpage

\bibliography{myBib}

\end{document}

%
%%%%%%%%%% COMMENT ABOVE THIS LINE FOR USE AS INCLUDE FROM MAIN TEXT

\setcounter{page}{1}
\setcounter{tocdepth}{1}
\renewcommand{\thefigure}{S\arabic{figure}}
\setcounter{figure}{0}
\renewcommand{\thetable}{S\Roman{table}}
\setcounter{table}{0}

%\renewcommand{\refname}{xx} % Was not needed since there was a manual heading

%FOR SDFT Section
\newcommand{\be}[1]{\begin{eqnarray}  {\label{#1}}}
\newcommand{\ee}{\end{eqnarray}}

\begin{center}
\textbf{{\LARGE Supplementary Information}}

for

\textbf{{\Large Spin-relaxation times exceeding seconds for color centers with strong spin-orbit coupling in SiC}}

by

Carmem~M.~Gilardoni *, Tom~Bosma *, Danny~van~Hien, Freddie~Hendriks, Bj\"orn Magnusson, Alexandre Ellison, Ivan G. Ivanov, N.~T.~Son and Caspar~H.~van~der~Wal

\emph{Version of \today}

\end{center}

\vspace{4cm}

\tableofcontents

\newpage

\noindent \textbf{References for Supplementary Information}
\bibliography{myBib}

%\newpage
%\noindent \textbf{TABLE OF CONTENTS}

%%%%%%%%%%%%%%%% END ARXIV TOP PART

\newpage

\section{Experimental methods}

\paragraph{Setup}

The laser power is $200~\mu W$ for the resonant pulsed beam at 1121.32~nm. It is polarized linearly along the 6H-SiC c-axis. The repump beam accounting for charge state switching has 2~mW power at 770~nm, it is pulsed with 80~MHz repetition rate.

\paragraph{Fitting routine}

We extract the $T_1$ lifetimes from the time-resolved PLE experiments in Fig.~$2$(c) (main text) by comparing the leading edge response from the first pulse $h_1$ to that from the second pulse $h_2$. These are determined by taking the average of the first 20 bins (of varying size, dependent on the timescale at a certain temperature) of the responses to the pulses. The behavior of the fraction of $h_2$ and $h_1$ can be described as
\begin{equation}
\frac{h_2}{h_1} = \frac{h_0(1-e^{-t/T_1})+\delta}{h_0 + \delta}
\end{equation}
where $\delta$ is the baseline height, $h_0$ is the peak height at thermal equilibrium minus the baseline $\delta$, and $t$ is the delay between both pulses. This allows us to create a fit function
\begin{equation}
\label{eq:fitfun}
\frac{h_2}{h_1} = q(1-e^{-t/T_1}) + 1-q
\end{equation}
with only two free parameters: $T_1$ and $q$, which represents the fraction $\frac{h_0}{h_0+\delta}$. 

\paragraph{Deadtime pile-up effect}

For the experiments in this work we used an avalanche photodiode single photon counter (SPC) for detection. After every detection event the detector has to reset its state, for which a predefined deadtime is used ($10~\mu$s in our case). This creates a pile-up effect in our experiments when integrating many pulse sequences. We investigated how this causes an error in determining $T_1$.

The probability of having a count at time $t$ can be described by the combined probability that a photon is present and detected at the counter $P_\text{photon}$ multiplied by the probability of not having measured a photon before for the duration of the deadtime $\Delta t$. We get the equation
\begin{equation}
P_\text{count}(t) = P_\text{photon}(t) \left(1 - \int_{t-\Delta t}^{t}P_\text{count}(\tau)d\tau\right)
\end{equation}

\begin{figure}[h!]
	\centering
	\includegraphics[width=7cm]{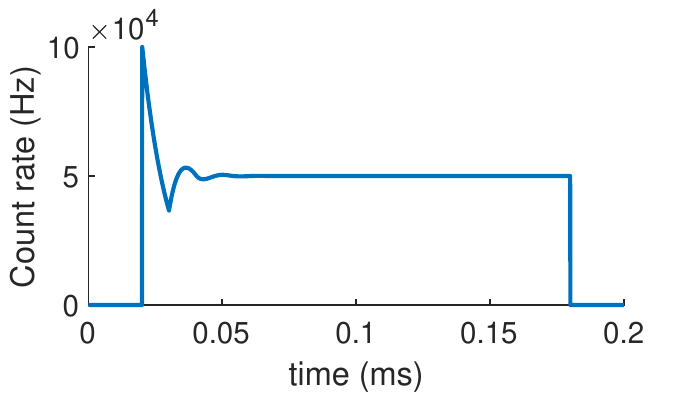}
	\caption{\textbf{}}
	\label{fig:Fig_Deadtime}
\end{figure}

The outcome is plotted in Fig.~\ref{fig:Fig_Deadtime} for a square PLE response to a pulse where $P_\text{photon}(t)=100~$kHz during the pulse, which saturates the SPC. Before the pulse, the detector is always ready to receive a photon, thus the initially measured count rate relates accurately to the number of photons present. For the duration of the deadtime ($10~\mu$s) the detector will be recovering and thus the count rate drops. This is periodic, but it clearly damps out to a steady-state count rate due to a spread in the actual detection times.

This will give a positive error $\epsilon_1$ ($\epsilon_2$) for the measured value of $h_1$ ($h_2$). At small $t$ the magnitude of the error $\epsilon_2$ is low and increases to $\epsilon_1$ as $t\gg T_1$. If we rewrite Eq.~\ref{eq:fitfun} to represent the measured values it becomes
\begin{equation}
\frac{h_2}{h_1} = q'(1-e^{-t/T_1}) + 1-q' + \frac{\epsilon_1 - \epsilon_2}{h_0 + \delta - \epsilon_1}
\end{equation} where $q'$ is represents the corrected fraction $\frac{h_0}{h_0+\delta-\epsilon_1}$. The final term in this equation makes the measured $\frac{h_2}{h_1}$ approach its asymptote faster than in reality. Thus, fitting $\frac{h_2}{h_1}$ to Eq.$~\ref{eq:fitfun}$ will always yield shorter $T_1$ values compared to the true spin-flip times. Simulating the pile-up effect for our measurements yields errors in $T_1$ varying from $2\%$ to $7\%$. We choose to not correct for this and report the lower-bound values for the $T_1$ spin-flip times. Background PLE and dark counts have been neglected in this analysis as we found them to be of little influence so long as they remain below 10\% of the peak count rate.
\clearpage

\section{Zero-field measurement}
\label{secSI:zerofield}

\begin{figure}[h!]
	\centering
	\includegraphics[width=7cm]{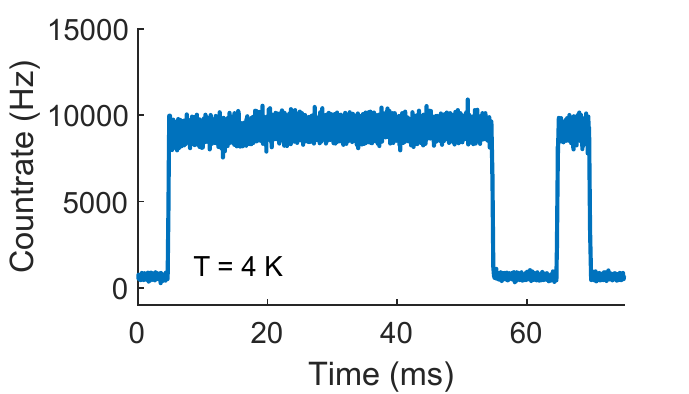}
	\caption{\textbf{Zero field measurement.} Time-resolved PLE measurement at $B=0$~T, $T = 4~$K and $200~\mu$W excitation power. For this particular experiment, the excitation pulses were right-circularly polarized, but we found no unexpected dependence on polarization: the PLE response was always proportional to the component of linear driving along the c-axis.}
	\label{fig:Fig_NoField}
\end{figure}

To confirm that the $T_1$ lifetimes measured are actually from the spin from the $\ket{G1}$ and not influenced by some other process, we performed zero-field experiments. Both ground-state spins are then degenerate and no spin pumping is expected. The results are shown in Fig.~\ref{fig:Fig_NoField}. The leading-edge peak in PLE as seen in Fig.~$2$(c) in the main text vanishes in Fig.~\ref{fig:Fig_NoField}(a). We note that thermal effects from laser driving at high intensities become more prominent at zero magnetic field, since darkening of the PLE no longer occurs. Additionally, we checked again the optical polarization dependence for these conditions at $B = 0$~mT, but this confirmed that the driving was still only sensitive to the component parallel to the c-axis.

\begin{figure}[h!]
	\centering
	\includegraphics[width=14cm]{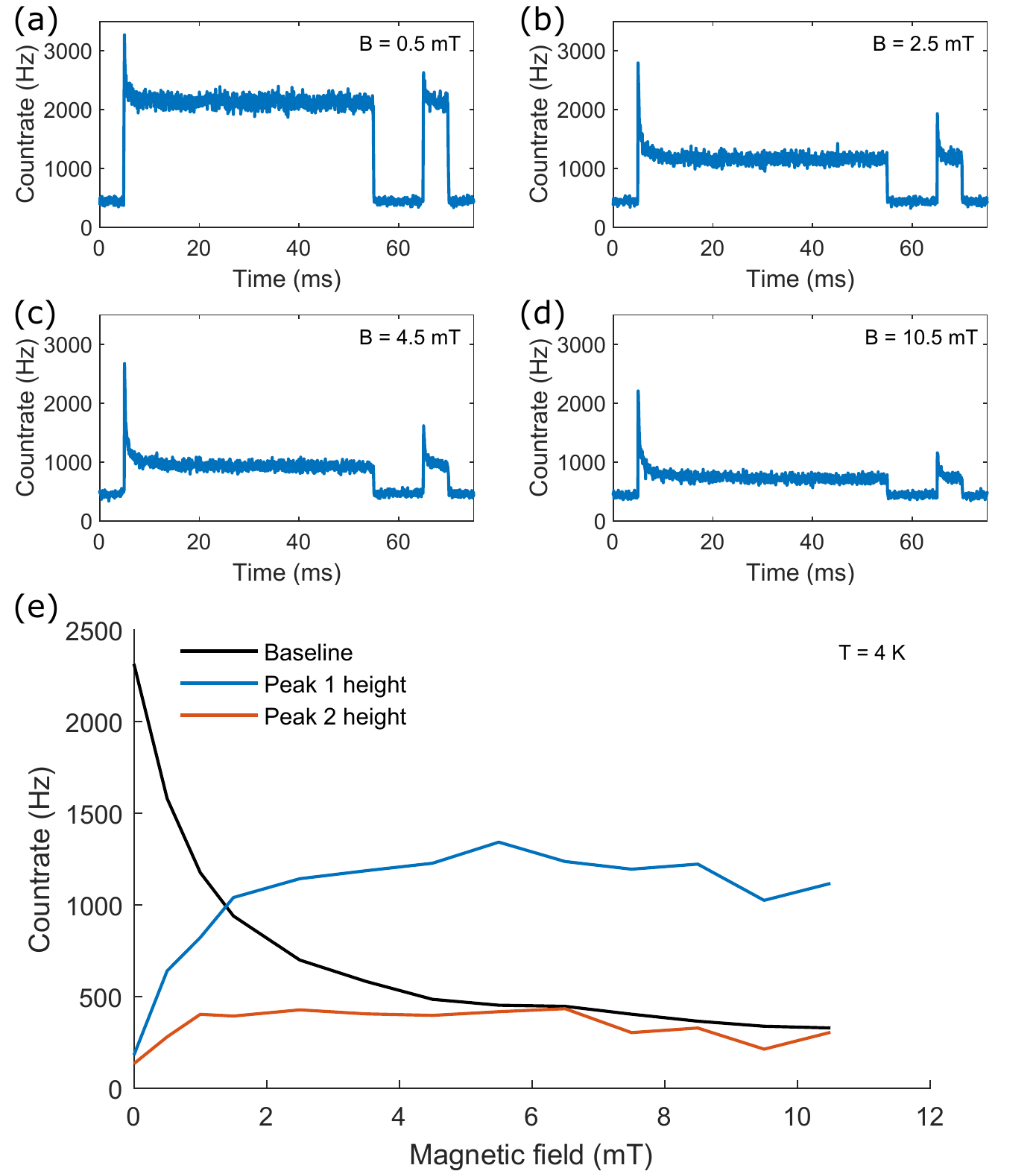}
	\caption{\textbf{Magnetic field dependence} (a-d) Time resolved measurements at various magnetic fields. (e) Baseline and leading-edge peak heights (baseline subtracted) versus magnetic field. A repump laser counteracting any bleaching was always present in these experiments. All measurements at $T = 4~$K.}
	\label{fig:Fig_Bsweep}
\end{figure}

In Fig.~\ref{fig:Fig_Bsweep} we show the evolution of the PLE response with magnetic field strength. Figure~\ref{fig:Fig_Bsweep}(e) summarizes this experiment, depicting the steady-state baseline of the PLE response together with the leading-edge peak heights (baseline subtracted) upon ramping the field. Up to 4 mT the baseline decreases, indicating that spin-pumping occurs to an increasing degree. Beyond $4$~mT it stabilizes. At this field the spin-splitting ($49$~MHz at $4$~mT) is well beyond the homogeneous linewidth ($15$~MHz \cite{bosma2018}) of the spin-states, and a single laser can no longer drive transitions from both spin states.

% Hom. linewidth = 15 MHz
% At 2 mT the splitting is 18 MHz, so outside the hom. linewidth and spin features saturate

% Zero-field measurements
% 280 uW scanning beam, 2.3 mW repump (770 nm)

% B-field sweep
% T = 4K, repump always on (770 nm, 3.3 mW)
% 200 uW scanning beam

% Data from '20190430 - T1 - ZeroFieldMeasurements' secondary data (20190503_n2)

% Data from '20190529 - T1 at first B-field sweep' secondary data

\clearpage
\section{Charge-state switching}

When resonantly addressing optical transitions in the Mo defect in SiC, the PLE response drops over time. This is ascribed this to charge-state switching of the defect. We investigate the timescales of this bleaching in order to rule out its influence on the measurements of the $T_1$ spin-flip times.

\begin{figure}[h!]
	\centering
	\includegraphics[width=14cm]{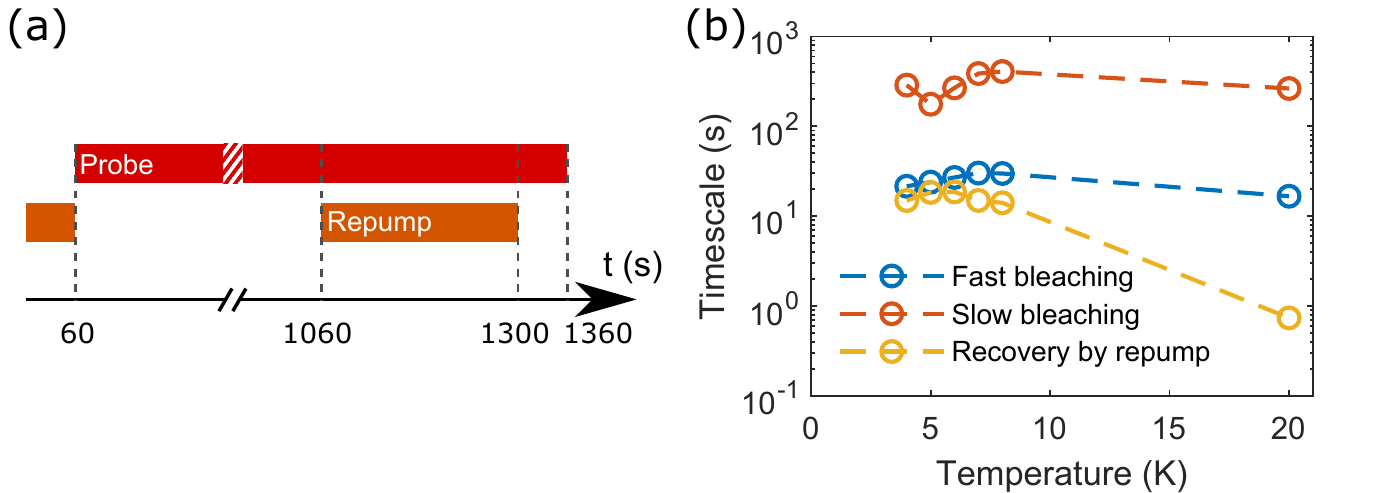}
	\caption{\textbf{Bleaching experiment} (a) Sequence of lasers used to measure the bleaching/repump timescales. (b) The acquired timescales versus temperature.}
	\label{fig:Fig_BleachingTimes}
\end{figure}

The experimental approach is depicted in Fig.~\ref{fig:Fig_BleachingTimes}(a). First, a repump beam ($770$~nm, pulsed) counteracts any prior bleaching \cite{bosma2018}, resetting the charge state for $60$ seconds. Next, a probe beam resonant with the ZPL illuminates the sample, slowly bleaching the Mo defects. We can track the bleaching timescale by measuring the PLE response as function of time. After another $1000$ seconds the repump beam is incident on the sample together with the probe beam, which allows us to track the recovery timescale. Finally, the repump is switched off for 60 seconds to check the initial decay of the PLE response. This sequence is repeated four times. We use a 2 mW repump beam and a $200~\mu$W probe focused to approximately $100~\mu$m diameter in the sample. The magnetic field strength is $100$~mT at an angle of $\ang{57}$ with the c-axis.

\begin{figure}[h!]
	\centering
	\includegraphics[width=14cm]{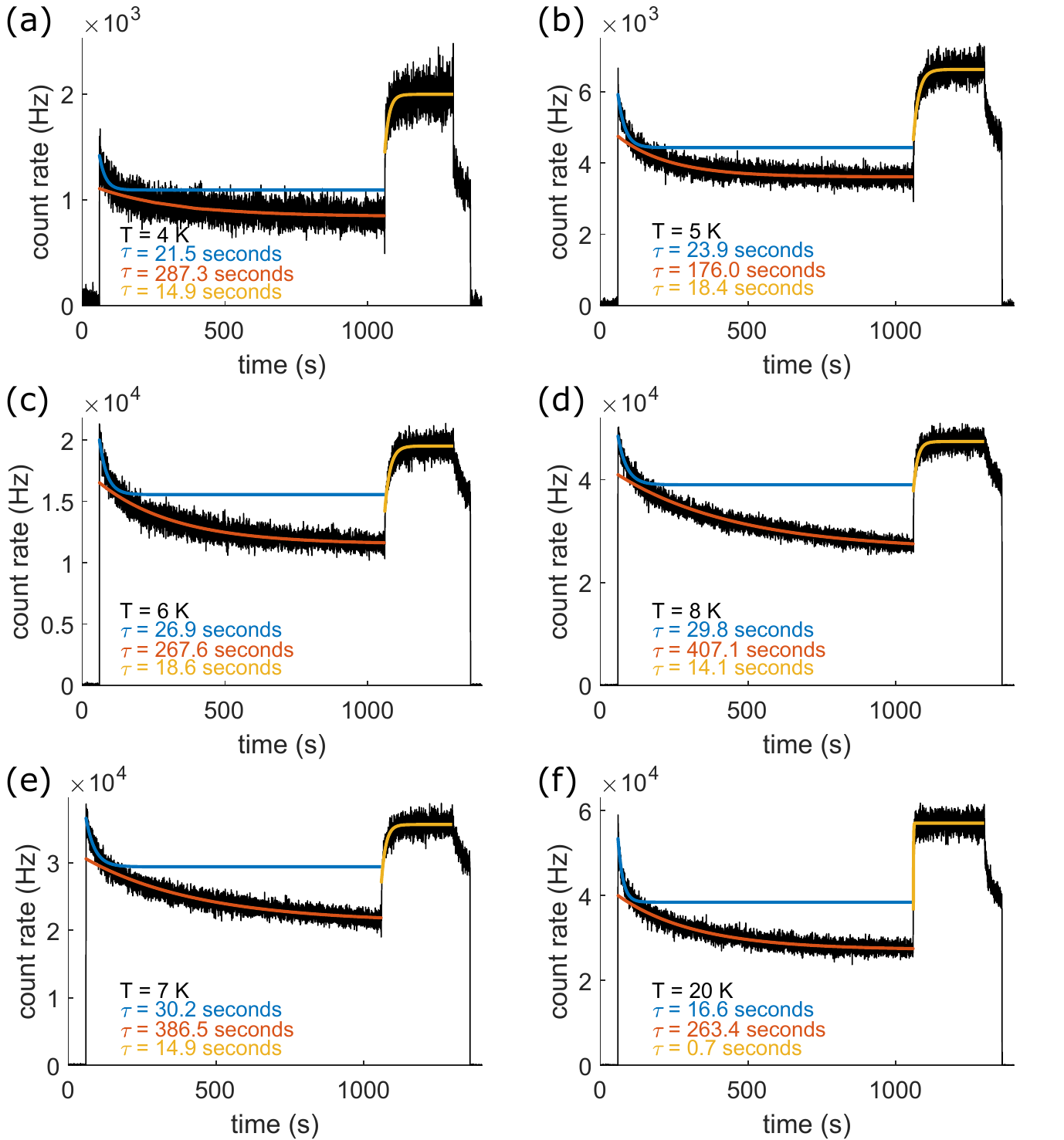}
	\caption{\textbf{Detailed bleaching results} (a-f) Dynamics of the resonant photoluminescence signal (PLE) over long time periods, at various temperatures. A static background is subtracted as well as a fluorescence background induced by the repump when it is on (measured during the first 60 seconds). When the repump is off, but the system is being renonantly driven, the PLE signal decays exponentially with two characteristics time constants (blue and red fits). As the repump is turned on, the PLE signal recovers quickly, with a single time constant (yellow fit).}
	\label{fig:Fig_Bleaching}
\end{figure}

The results are shown in Fig.~\ref{fig:Fig_Bleaching}. The bleaching between 60 and 1060 seconds occurs according to two timescales, both are fit with an exponential decay for the orange and blue curve. The yellow curve is an exponential fit to the recovery by the repump laser. All three timescales are plotted versus temperature in Fig.~\ref{fig:Fig_BleachingTimes}(b). For the $T_1$ experiments, where the repump beam was only on in between measurement runs, the two bleaching scales are most relevant. Both occur at rates that are at least one order of magnitude slower than the observed spin-flip times. Any bleaching occurring at faster timescales should have been visible in the zero-field measurements from section \ref{secSI:zerofield}. Thus, the effect of bleaching on measuring $T_1$ can be deemed negligible. 

Note the fast decay of PLE for $4$~K (Fig.~\ref{fig:Fig_Bleaching}(a)) after the repump laser is blocked at $1300$ seconds. We ascribe this to fast spin flips induced by the repump laser. When performing the experiments to measure $T_1$ with an omnipresent repump laser, we observe that $T_1$ is reduced by an order of magnitude (for similar laser powers). This fast decay is not visible in Fig.~\ref{fig:Fig_Bleaching}(b-f) since the spin-flip times are already quite short at higher temperatures.
% We should measure this at zero field...

\clearpage

\section{Electronic structure - group theoretical approach} \label{sec:GroupTheory}

The Mo defect is characterized by a single active spin in the $4$d shell of a molybdenum impurity at a Si substitutional lattice site \cite{bosma2018}. In this configuration, the Hamiltonian of the defect has a three-fold rotational symmetry and three vertical mirror planes, such that the eigenfunctions of the electronic orbital state transform according to the symmetry group C\textsubscript{$3$v}. Despite the extensive literature available on the effect of crystal field and spin-orbit coupling on the electronic states of these defects \cite{kaufmann1997,kunzer1993}, we are unaware of a comprehensive report on the effect of symmetry on the coupling terms between the various specific spin sublevels and external fields based on the double group representations of the defect symmetry, and we present this analysis here. We apply a group-theoretical approach to obtain the symmetry of the eigenfunctions associated with a defect center in a crystal field of C\textsubscript{$3$v} symmetry in the presence of spin-orbit coupling. Furthermore, we obtain and explain the selection rules governing the interaction of the electronic spin with magnetic and electric fields. 

%%%%%%%%%%%%%%%%%%%%%%%%%%%%%%%%%%%%%%%%
%%%%%%% Point group of defect, including SOC - Figure: defect structure, Front-Top view
%%%%%%%%%%%%%%%%%%%%%%%%%%%%%%%%%%%%%%%%

The group theoretical rules governing the selection rules presented in the end of this work do not rely on a particular basis set for the description of the electronic wavefunction. By this, we mean that even if we consider hybridization of the wavefunction of the bare transition metal atom/ion with the nearest carbon atoms due to covalent bonding, the symmetry of the crystal field Hamiltonian is preserved such that the new, modified wavefunctions will still obey the selection rules arising from a group-theoretical analysis. Nonetheless, it is instructive to start from an analysis of the effect of the Hamiltonian on the $10$ spin-orbital states arising from a single electron sitting in one of the d-orbitals of the transition metal.

\begin{figure}[h!]
	%\centering
	\includegraphics[width=10cm]{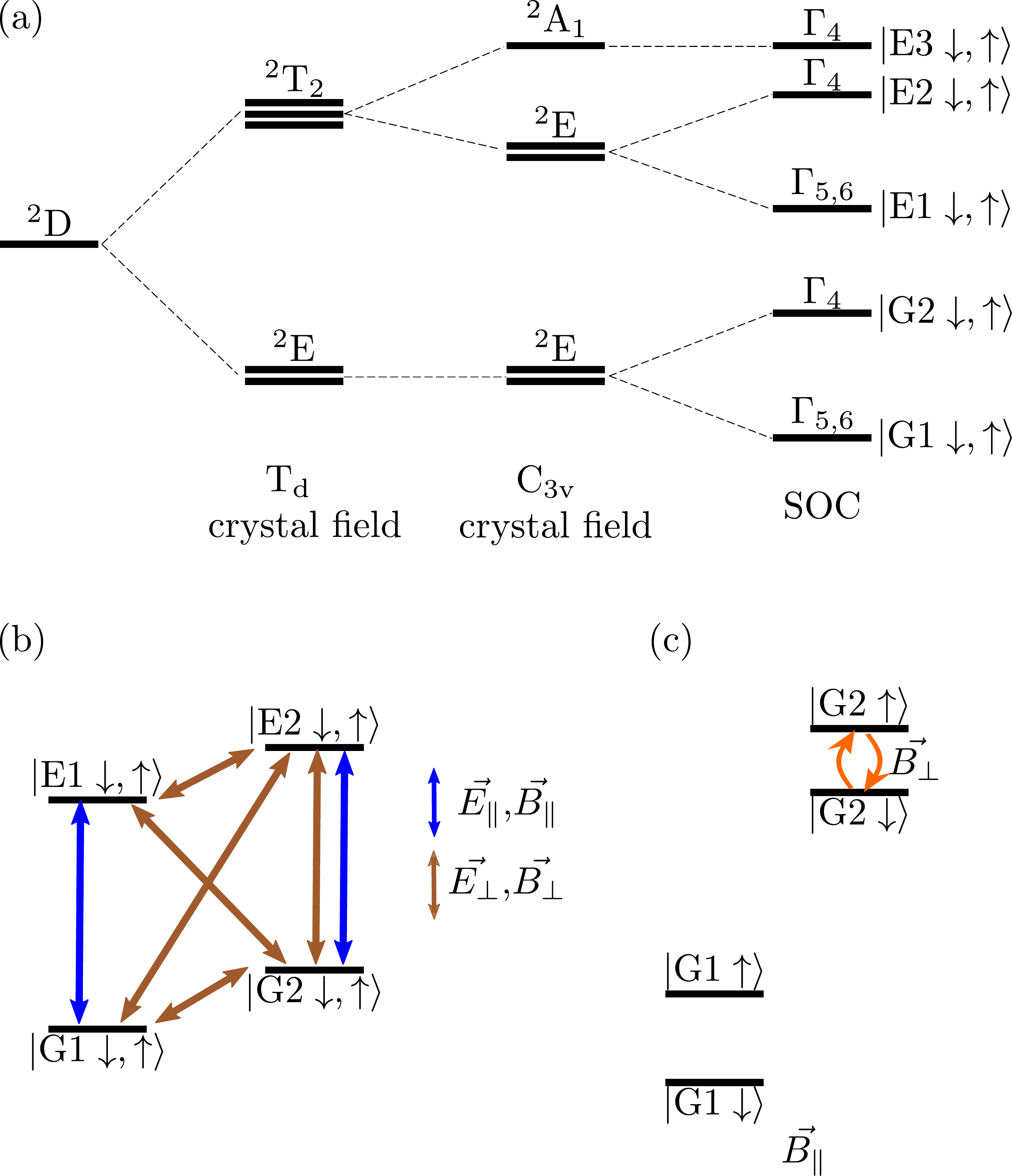}\\
	\caption{\textbf{Electronic Structure and selection rules based on group-theoretical analysis} (a) The $10$ spin-orbital states corresponding to the $^2$D configuration of the free ion are split into $5$ Kramer's doublets under the action of the crystal field and SOC, of which three transform as irrep $\Gamma_4$ and two transform as irrep $\Gamma_{5,6}$ of the double group \=C\textsubscript{$3$v} (see main text). (b) Coupling between $2$ KD transforming as $\Gamma_{5,6}$ is allowed under the action of an electric or magnetic field parallel to the symmetry axis; coupling between $2$ KD transforming as $\Gamma_4$ is allowed under the action of an electric or magnetic field pointing in any direction; coupling between a KD transforming as $\Gamma_{5,6}$ and a KD transforming as $\Gamma_4$ is allowed under the action of an electric or magnetic field perpendicular to the symmetry axis. (c) A KD transforming as $\Gamma_{5,6}$ does not interact with a magnetic field perpendicular to the symmetry axis, whereas this is not the case for a KD transforming as $\Gamma_4$.}
	\label{fig:FigSI1}
\end{figure}

The transition metal at a silicon substitutional site shows tetrahedral coordination due to bonding to the $4$ nearest carbons. In this configuration, the $5$ d orbitals split into an orbital doublet and an orbital triplet (which transform as the irreps E and T\textsubscript{$2$} of the symmetry group T\textsubscript{d}), where the triplet lies highest in energy. Due to the hexagonal character of the lattice, the tetrahedral symmetry of these sites is lowered to C\textsubscript{$3$v}, with the rotational axis aligned parallel to the growth axis of the crystal. Upon this symmetry reduction, the triplet further splits giving rise to an orbital doublet and a singlet, which transform respectively as E and A\textsubscript{2}. The effect of this symmetry lowering operation is expected to be largest in the lattice sites of quasi-hexagonal symmetry ($h$), and to only modestly affect the lattice sites of quasi-cubic symmetry ($k$) (Fig.~\ref{fig:FigSI1}(a)).

A wavefunction transforming as a \emph{non-degenerate} irrep of a given point-group cannot have an effective orbital angular momentum (in other words, the orbital angular momentum is quenched) \cite{abragam1970}. However, this requirement is lifted in the presence of degeneracies, such that the eigenfunctions of the Hamiltonian transforming as E are allowed to have a non-zero orbital angular momentum. Thus, in order to fully describe our system, we must consider the effect of spin-orbit coupling. In a group-theoretical approach, this is done by extending the group of interest to include $2\pi$ rotations which bring a spin $\uparrow$ into a spin $\downarrow$ \cite{dresselhaus2008}. That is, this is done by considering the eigenfunctions as basis states of the irreps of the double group associated with the C\textsubscript{$3$v} group, here denoted by \=C\textsubscript{$3$v}.

In the double group including the effect of spin-orbit coupling, three irreps describing how odd spin wavefunctions transform are added to the group. These irreps are $\Gamma_4$, which is doubly degenerate, and $\Gamma_{5,6}$, two irreps that are connected by time-reversal symmetry and must thus be degenerate in the presence of time-reversal symmetry. The orbital singlet transforming as A\textsubscript{$2$} gives rise to a Kramer's doublet (KD) transforming as $\Gamma_4$, whereas an orbital doublet transforming as E splits into two KDs, of which one transforms as irrep $\Gamma_4$, and the other transforms as irreps $\Gamma_{5,6}$. Thus, the symmetries mentioned above split the $10$ states arising from an electronic configuration \textsuperscript{$2$}D into $5$ Kramer's doublets, of which $2$ transform as $\Gamma_{5,6}$, and $3$ transform as $\Gamma_4$ (Fig.~\ref{fig:FigSI1}(a)). The character table of the double group \=C\textsubscript{$3$v} is given in Tab. \ref{tab:CharacterDouble}. Additionally, we explicitly show what are the transformation properties of the vectors $x,y,z$ and the axial vectors $R_x,R_y,R_z$, as well as how the cubic harmonics $z^2, x^2-y^2, xy, xz, yz$ transform under the operations of the group.
 
We can investigate the role of small magnetic and electric fields in driving transitions between different KD (coupling between different KDs), and spin resonances (coupling between the two eigenstates pertaining to a single KD) in the framework of group-theory, given that these fields are small enough that the symmetries of the Hamiltonian $H_0$ are preserved. The selection rules between two wavefunctions can be obtained in a straight-forward way. If $\ket{\psi_i}$ and $\ket{\phi_{i'}}$ are two eigenstates of the Hamiltonian $H_0$ transforming respectively as irreps $\Gamma_i$, $\Gamma_{i'}$, the selection rules with respect to a perturbative Hamiltonian $H'$ are given by the product $\bra{\psi_i}H'\ket{\phi_{i'}}$. In order for this matrix element to be non-zero, it must transform as a scalar, that is, as the totally symmetric irrep A\textsubscript{$1$} \cite{dresselhaus2008}. Thus, the product of the representations $\Gamma_i^* \otimes \Gamma_j \otimes \Gamma_{i'}$, where the perturbation $H'$ transforms as $\Gamma_j$ and $^*$ denotes complex conjugation, must contain the totally symmetric irrep A\textsubscript{$1$}. 

Table \ref{tab:Productirr} gives the decomposition of the various products of $\Gamma_4$, $\Gamma_{5,6}$, in terms of irreps of the C\textsubscript{$3$v} group, and translates this into the selection rules governing the coupling between various spin states. 

Optical transitions between various sets of KDs are allowed due to coupling to $\vec{E_\parallel}$, $\vec{E_\bot}$, which belong to irreps $\text{A}_1$ and $\text{E}$, respectively. We can extract polarization selection rules from table \ref{tab:Productirr}. Electric field driven transitions between two KDs transforming as $\Gamma_{5,6}$ will be polarized along the symmetry axis of the defect; transitions between two KDs transforming as $\Gamma_4$ can be polarized along any direction; transitions between a KD transforming as $\Gamma_{5,6}$ and a KD transforming as $\Gamma_4$ are only allowed for light polarized perpendicular to the symmetry axis. These properties are summarized in Fig.~\ref{fig:FigSI1}(b). This means that only electric or magnetic fields of $\text{E}$ symmetry (that is, in the $xy$ plane) are capable of coupling and mixing states $\ket{G1}$ (which transforms as $\Gamma_{5,6}$) and $\ket{G2}$ (which transforms as $\Gamma_{4}$).

Transitions and energy splittings within each of the KDs can also be understood based on the symmetry of the defect. The anisotropic Zeeman structure observed for the ground state spin doublet, which is insensitive to magnetic fields perpendicular to the crystal symmetry axis \cite{bosma2018} can be understood purely based on the properties of the group. A magnetic field along the symmetry axis of the defect transforms as $R_z$, whereas a magnetic field perpendicular to this axis transforms as $R_x,R_y$. Within a doublet which transforms as $\Gamma_{5,6}$, no coupling is allowed with a magnetic field perpendicular to the symmetry axis since $\Gamma_{5,6}^* \otimes \text{E} \otimes \Gamma_{5,6} = \text{E} \not\supset \text{A}_1$. This is not the case for a doublet transforming as $\Gamma_4$, such that the spin sublevels that transform as $\Gamma_4$ are allowed to couple to magnetic fields in the plane, and will not have $g_\bot = 0$. Thus, we conclude that the ground state doublet belongs to the irrep $\Gamma_{5,6}$ (Fig.~\ref{fig:FigSI1}(c)). As long as the quantization axis of the defect spin (states $\ket{G1}$, belonging to irrep $\Gamma_{5,6}$) points parallel to the symmetry axis of the defect, we cannot rotate the spin via microwave spin resonances, since these spins are insensitive to magnetic or electric fields perpendicular to this axis. 

Finally, we note that if two spin sublevels are strictly connected by time-reversal symmetry (that is, they are a pure KD), they cannot be connected by operators that preserve time-reversal symmetry. This was proven by Kramer and became what is known as Kramer's theorem \cite{dresselhaus2008}. Thus, within a pure KD, electric fields are not capable of driving transitions between the two spin-sublevels.

%--------------------------
%------- Table 1 ----------
%--------------------------

\begin{table}
	\caption{\textbf{Character table of the double group \=C\textsubscript{$3$v}}. The upper bar represents an operation followed by a $2\pi$ rotation that brings a spin $\uparrow$ into $\downarrow$.}
	\label{tab:CharacterDouble}
	\begin{center}
		\begin{tabular}{ l | l r r r r r | c | c}
			\hline
			& E & \=E & $2\text{C}_{3}$ & $2\text{\=C}_{3}$ & $3\sigma_v$ & $3\bar{\sigma}_v$ & & \\
			\hline
			$\text{A}_1$ & $1$ & $1$ & $1$ & $1$ & $1$ & $1$ & $z$ & $z^2, x^2 + y^2$\\
			$\text{A}_2$ & $1$ & $1$ & $1$ & $1$ & $-1$ & $-1$ & $R_z$ & \\
			$\text{E}$ & $2$ & $2$ & $-1$ & $-1$ & $0$ & $0$ & $(x,y), (R_x,R_y)$ & $(x^2 - y^2, 2xy), (xz,yz)$\\ \hline
			$\Gamma_4$ & $2$ & $-2$ & $1$ & $-1$ & $0$ & $0$ & & \\
			\multirow{2}{*}{$\Gamma_{5,6}$\Bigg\{ } & $1$ & $-1$ & $-1$ & $1$ & $i$ & $-i$ & & \\
			& $1$ & $-1$ & $-1$ & $1$ & $-i$ & $i$ & &\\
			\hline
		\end{tabular}
	\end{center}
\end{table}

%--------------------------
%------- Table 2 ----------
%--------------------------

\begin{table}
	\caption{\textbf{Product tables of the double irreps of \=C\textsubscript{$3$v} and corresponding selection rules.}}
	\label{tab:Productirr}
	\begin{center}
		\begin{tabular}{ c | c | c c}
			\hline
			& $\Gamma_4$ & $\Gamma_5$ & $\Gamma_6$ \\
			\hline
			$\Gamma_4^*$ & $\text{A}_1 + \text{A}_2 + \text{E}$ & $\text{E}$ & $\text{E}$\\
			& $\vec{E_\parallel}$, $\vec{B_\parallel}$, $\vec{E_\bot}$, $\vec{B_\bot}$ & $\vec{E_\bot}$, $\vec{B_\bot}$ & $\vec{E_\bot}$, $\vec{B_\bot}$ \\ \hline
			$\Gamma_5^* = \Gamma_6$ & $\text{E}$ & $\text{A}_1$ & $\text{A}_2$ \\
			& $\vec{\text{E}_\bot}$, $\vec{B_\bot}$ & $\vec{E_\parallel}$ & $\vec{B_\parallel}$\\
			$\Gamma_6^* = \Gamma_5$ & $\text{E}$ & $\text{A}_2$ & $\text{A}_1$ \\
			& $\vec{E_\bot}$, $\vec{B_\bot}$ & $\vec{B_\parallel}$ & $\vec{E_\parallel}$\\
			\hline
		\end{tabular}
	\end{center}
\end{table}
\clearpage

\section{Simulation of raw data}

Due to the large number of available states for defect (vibrational levels, orbital state $\ket{G2}$ in Fig.$1$ of the main text, ionized states), it is not straight forward to obtain quantitative information from the shape of the raw data plots presented in the main text. Nonetheless, we can apply a rate equation model to reproduce the data and, upon carefully taken assumptions, obtain a bound for the values of the optical decay time and Rabi driving frequency in our experiments.

\begin{figure}[h!]
	%\centering
	\includegraphics[width=8cm]{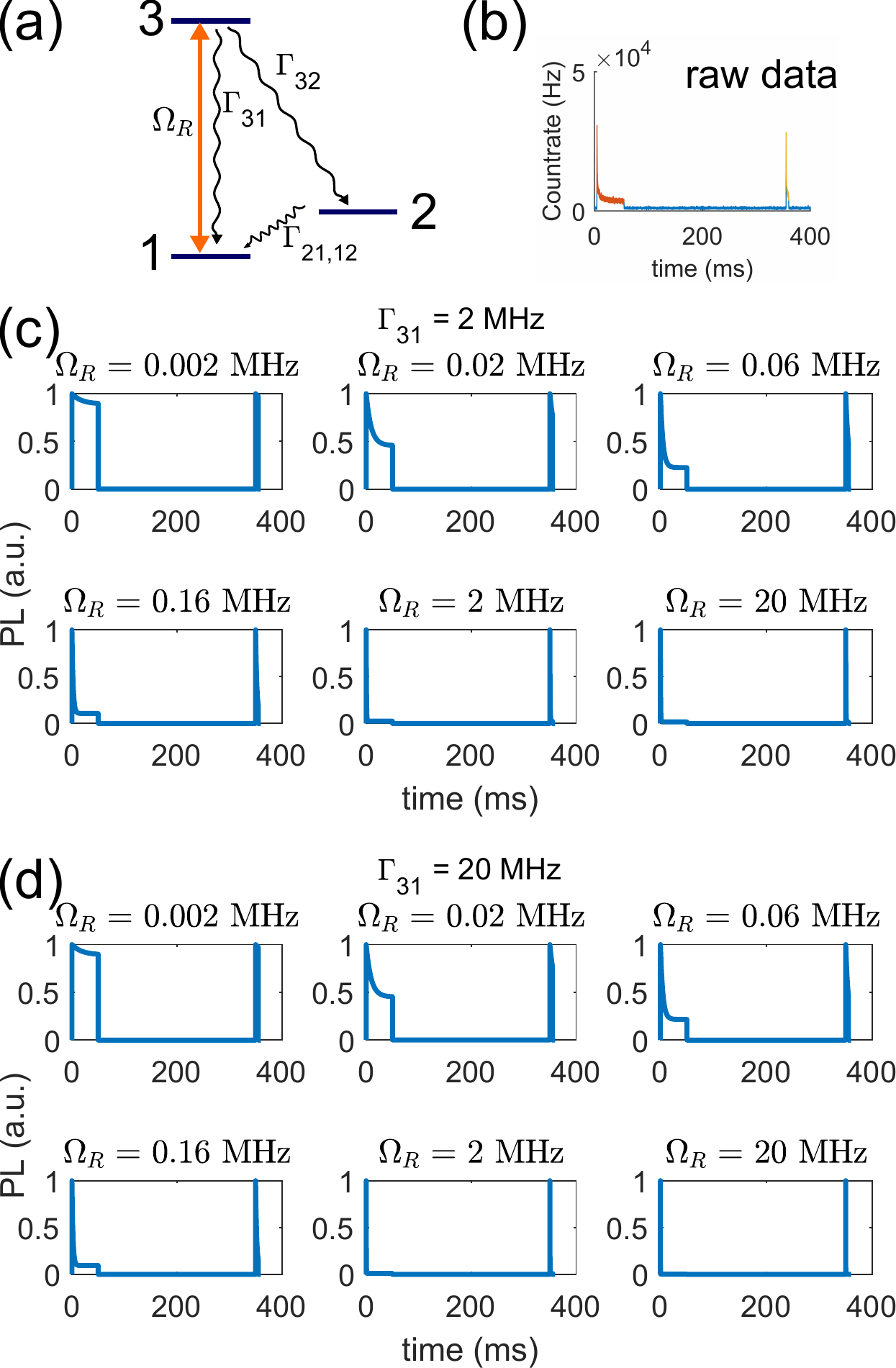}\\
	\caption{\textbf{Three-level model describing dynamics of defect in ms timescale} (a) The three level model inclued a ground and optically excited states that are connected via optical driving with Rabi frequency $\Omega_R$. The optically excited state can decay into the ground state or into a shelving state, with rates $\Gamma_{31}$ and $\Gamma_{32}$. Population can decay back from the shelving state into the ground state with a rate $\Gamma_{21}$. The photoluminescence excitation signal obtained is assumed to be proportional to the population in the optically excited state. (b) Typical experimental data in the presence of a magnetic field consists of PLE signal with sharp on and offset as the laser turns on and off. After a leading edge peak, the PLE signal decays back to a constant non-zero value, indicative of PL darkening due to optical pumping into shelving states within the first few microseconds of illumination. (c,d) Simulating the dynamics of the population in the excited state upon illumination and comparing these curves to the experimental data, we can determine that the Rabi frequencies observed in these experiments are of the order of a few tens of kHz.}
	\label{fig:Rate}
\end{figure}

In order to minimize the set of free parameters and facilitate the analysis of the behavior of the system, we simulate this defect center as a three-level system (Fig.~\ref{fig:Rate}(a)), where the ground (state $1$) and excited (state $3$) states can be coupled by an optical field with Raby frequency $\Omega_R$. From the optically excited state, the system can decay either back into the ground state with a rate $\Gamma_{31}$, or into a shelving state $2$ with a rate $\Gamma_{32}$. Additionally, population can be transferred between states $2$ and $1$ at a rate $\Gamma_{21}$, and between $1$ and $2$ at a rate $\Gamma_{12} = e^{-\Delta/kT} \Gamma_{21}$, where $\Delta$ is the energy difference between $2$ and $1$ and $kT$ denotes the thermal energy of the system. We simulate the system with a simple set of rate equations for the populations of each state ($P_1, P_2, P_3$), without treating coherences explicitly. Finally, we consider that the photoluminescence observed is proportional to the population of the optically excited state, $P_3$. 

We try to reproduce the typical shape of the raw PL data obtained experimentally (Fig.~\ref{fig:Rate}(b)) in order to obtain a set of reasonable values for the Rabi driving frequency and optical decay rates in our system. We assume that states $1$ and $2$ correspond to the ground state spin sublevels, $\ket{G1_\downarrow}$ and $\ket{G1_\uparrow}$ respectively. In this way, we can write $\Gamma_{32}$ in terms of $\Gamma_{31}$ by assuming that the branching ratios correspond to the overlap of the spin states in ground and optically excited state \cite{bosma2018}. This gives $\Gamma_{32} \sim 0.003 \Gamma_{31}$. Previous experiments revealed an excited state lifetime of $\sim 56$~ns, resulting in $\Gamma_{31} \sim 20$~MHz. Similar TM defects have been recently reported with optical excited state lifetime of $\sim 100$~ns. Thus, we simulate the experiment for values of $\Gamma_{31}$ of $2$ and $20$~MHz. For $\Gamma_{21}$, we use the values presented in the main text for the spin lifetime (Fig.~$3$).

We note that differences based on the exact value of $\Gamma_{31}$ are barely noticeable. Thus, we cannot restict our estimate for $\Gamma_{31}$ further. Nonetheless, we can restrict the expected values of $\Omega_R$ by comparing the calculated traces presented in Fig.~\ref{fig:Rate}(c,d) and the raw trace presented in Fig.~\ref{fig:Rate}(b). We not that if $\Omega_R$ is very small, of the order of a kHz, PL darkening is almost absent, unlike what is seen in experiment, where PL darkening is significant. In contrast, if $\Omega_R$ is of the order of a few MHz, the defect darkens completely within the time of the driving pulse. This is also in disagreement with the experimental data. Thus, we conclude that the Rabi frequency in our experiements is of a few tens to hundreds of kHz.  

\begin{figure}[h!]
	%\centering
	\includegraphics[width=10cm]{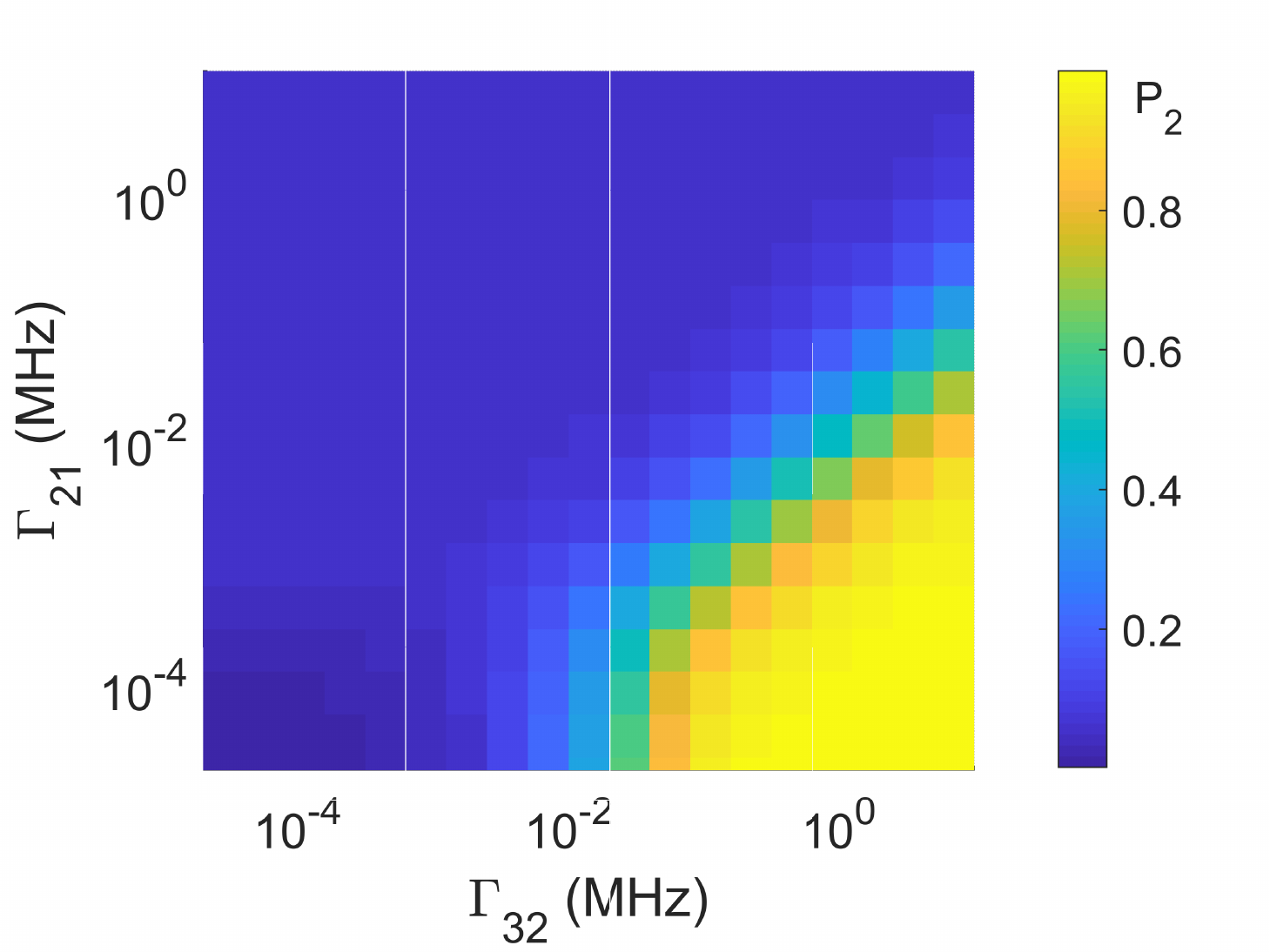}\\
	\caption{\textbf{Optical pumping into orbital state $\ket{G2}$} Significant population is only transferred into $\ket{G2}$ when $\Gamma_{32} \gg \Gamma_{21}$.}
	\label{fig:OpticalPumping}
\end{figure}

Furthermore, section~\ref{secSI:zerofield} shows that we do not see any PL darkening when we perform the time-resolved measurements described in the main text at zero magnetic field. In this case, state $2$ in our model corresponds to the orbital state $\ket{G2}$ from the main text. We calculate the population in state 2 after optically driving the system for approximately $500$~ms, with Rabi frequencies of the order of a few tens of kHz, and present these results in Fig.~\ref{fig:OpticalPumping}. We only transfer significant population into $2$ (leading to PL darkening) when the optical decay rate into state $2$, $\Gamma_{32}$ is larger than the rate at which the system leaves state $2$, $\Gamma_{21}$. Since we do not observe any PL darkening, we conclude that $\Gamma_{32} \ll \Gamma_{21}$ such that, within the time of our measurements, no significant population is transferred into the orbital state $\ket{G_2}$, and the presence of this state does not influence our measured value for the spin $T_1$.

\clearpage 

\section{$T_1$ vs Temperature}

\begin{figure}[h!]
	\centering
	\includegraphics[width=14cm]{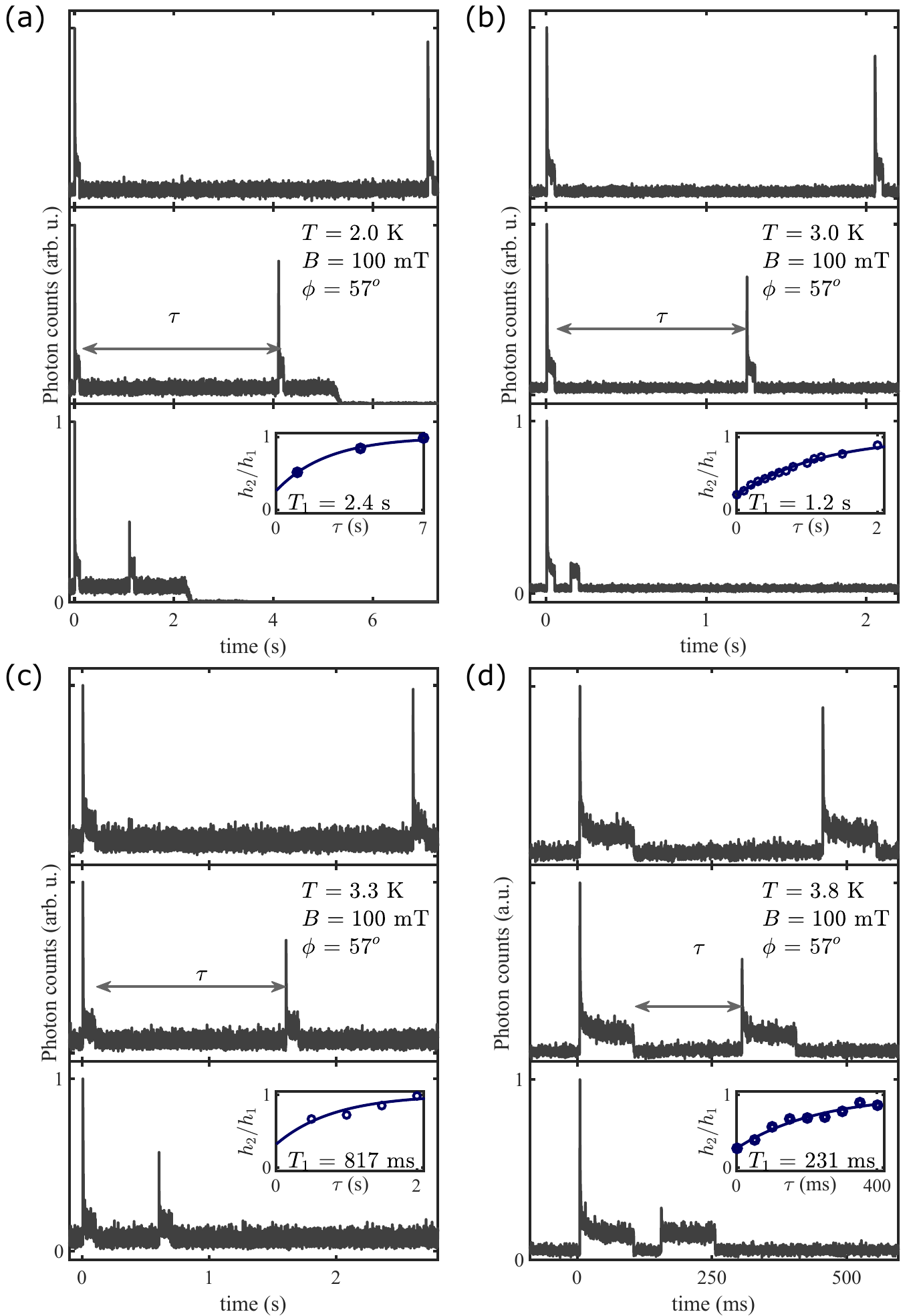}
\end{figure}

\begin{figure}[h!]
	\centering
	\includegraphics[width=14cm]{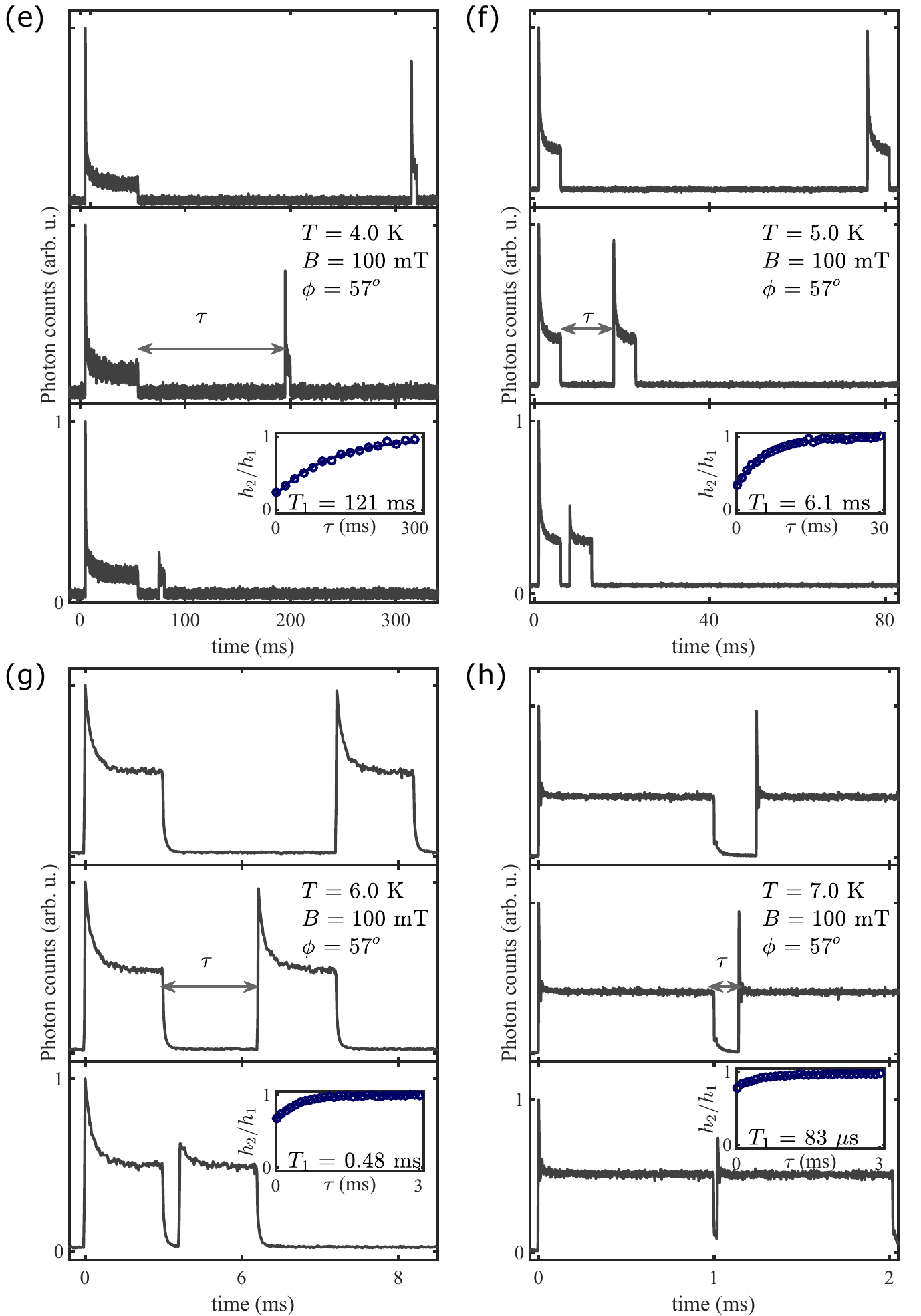}
	\caption{\textbf{Spin-flip times at various temperatures.} For the 2~K measurement (a) a shutter closed before and after the pulse train. The SPC data was only considered when the shutter was open and therefore the signal drops to zero shortly after the pulse.}
	\label{fig:Fig_allTemps}
\end{figure}

\clearpage 

\section{Fitting of $T_1^{-1}$ vs temperature}

The spin-lattice relaxation of single spins of substitutional defects in solid state materials arises from a modulation of the crystal field potential in time due to the presence of phonons, which perturbs the stationary crystal field ($V^{(0)}$) and couples various eigenstates of the time-independent Hamiltonian to each other \cite{abragam1970}. Thus, the probability of a spin flip to occur depends largely on the matrix elements of the time-dependent crystal field $V^{(1)}$ between the various electronic levels accessible to the defect. We thus define the terms $V_{orb}$ and $V_{vib}$, which indicate the order of magnitude of the matrix elements of $V^{(1)}$ connecting $\{\ket{G1\downarrow},\ket{G1\uparrow}\}$ to $\{\ket{G2\downarrow},\ket{G2\uparrow}\}$ and $\{\ket{G1_{vib}\downarrow},\ket{G1_{vib}\uparrow}\}$ respectively (see Fig.$1$ in main text for definitions).

The direct process, a one phonon interaction driving transitions between states $\ket{G1\downarrow}$ and $\ket{G1\uparrow}$ directly, is expected to show a temperature dependence of the kind $T_1^{-1} \propto (\hbar \omega)^2 |\bra{G1\downarrow} V^{(1)} \ket{G1\uparrow}|^2  T$ when $k_B T \gg \hbar \omega$, where $\hbar \omega$ is the Zeeman splitting between the spin sublevels $\ket{G1\downarrow}$ and $\ket{G1\uparrow}$. Since the states $\ket{G1\downarrow}$ and $\ket{G1\uparrow}$ are each other's time-reversal pair and $V^{(1)}$ preserves time-reversal symmetry, the matrix elements $|\bra{G1\downarrow} V^{(1)} \ket{G1\uparrow}|$ are identically zero (see section~\ref{sec:GroupTheory}). Nonetheless, the presence of a magnetic field or hyperfine interaction perturbs states $\ket{G1\downarrow}$ and $\ket{G1\uparrow}$ by mixing in states higher in energy, in such a way that we expect the direct process to be present with a magnitude roughly proportional to $ (\hbar \omega)^4 \frac{|V^{(1)}|^2}{\Delta^2} T$, where $V^{(1)}$ is now the matrix element of the time-dependent crystal field coupling states $\ket{G1\downarrow}$, $\ket{G1\uparrow}$ to a generic excited state $\ket{E}$ lying an energy $\Delta$ above $\ket{G1\downarrow}$, $\ket{G1\uparrow}$. All of the excited states shown in Fig.~$1$(b) are expected to contribute to this process, such that mixing with both the higher KD ($\ket{G2\downarrow}$, $\ket{G2\uparrow}$) and the vibronic states ($\ket{G1_{vib}\downarrow}$, $\ket{G1_{vib}\uparrow}$) should be considered. In this way, we expect a dependence of the kind  $T_1^{-1} \propto (\hbar \omega)^4 (\frac{|V_{orb}|^2}{\Delta_{orb}^2} + \frac{|V_{vib}|^2}{\Delta_{vib}^2})T$.

Additionally, a spin polarization in the defect can decay back to its equilibrium value via two-phonon processes comprising transitions into real (Orbach process) or virtual (Raman process) excited states. The former gives rise to an exponential temperature dependence of the type $T_1^{-1} \propto |\bra{G1\downarrow} V^{(1)} \ket{E} \bra{E} V^{(1)} \ket{G1\uparrow}| \Delta^3  \exp(-\Delta/k_B T)$ in the limit of $\Delta \gg k_B T$, where $\Delta$ is the energy difference between a generic excited state $\ket{E}$ and the KD $\ket{G1\downarrow}$, $\ket{G1\uparrow}$. Orbach processes relative to transitions into states $\ket{G2\downarrow}$, $\ket{G2\uparrow}$ are expected to give rise to a strong temperature dependence at temperatures below $1$ K and saturate at higher temperatures, when $\Delta_{orb} \sim k_B T$, and its exponential behavior is thus not visible in our data. In contrast, Orbach processes relative to transitions into vibronic levels are expected to contribute significantly to the temperature dependence of $T_1^{-1}$ at a few K, since $\Delta_{vib} \sim 10$ meV $\gg k_B T$ between $2$ and $8$~K (Fig.~$1$(d)) .Thus, we expect the Orbach process to give rise to a temperature dependence of the kind  $T_1^{-1} \propto |V_{vib}|^2 \Delta_{vib}^3  \exp(-\Delta_{vib}/k_B T)$.

Finally, second order Raman processes give rise to a temperature dependence of the kind $T_1^{-1} \propto |V^{(1)}|^4 T^5$ when $\Delta \ll k_B T$, or $T_1^{-1} \propto (\frac{|V^{(1)}|}{\Delta})^4 T^9$ when $\Delta \gg k_B T$, where $\Delta$ is the energy difference between levels $\{\ket{G1\downarrow}$, $\ket{G1\uparrow}\}$ and a generic level $\ket{E}$ which is coupled to $\ket{G1\downarrow}$, $\ket{G1\uparrow}$ via $V^{(1)}$. Thus, since $\Delta_{orb} \sim k_B T$, Raman processes involving states $\ket{G2\downarrow}$, $\ket{G2\uparrow}$ are expected to show a $ |V_{orb}|^4 T^5$ dependence, whereas Raman processes involving virtual transitions into the vibronic levels $\ket{G1_{vib}\downarrow}$, $\ket{G1_{vib}\uparrow}$ are expected to contribute a term $(\frac{|V_{vib}|}{\Delta_{vib}})^4 T^9$ to $T_1^{-1}$.

%%%%%%%%%Data%%%%%%%%%%%%%%%%%%%%%%%%%%%%% 

We fit the datain Fig.~$3$ of the main text to a model of the type $T_1^{-1} = C_{D} T + C_{R} T^{n} + C_{O} \exp(-\Delta/k_B T) + \Gamma_0$, where $C_{D,R,O}$, $\Delta$ and $\Gamma_0$ are fitting parameters, and $n = 5,9$. The parameter $\Gamma_0$ is included to account for temperature independent processes of spin relaxation. The fit quality does not improve significantly if we consider $n=9$ instead of $n=5$ for the Raman process involved. Thus, we are unable to determine which levels are involved in the Raman transitions responsible for spin relaxation between $3$ and $4$~K. In either case, however, the exponential increase of the spin relaxation rate above $4$~K is accounted for by an Orbach process where two phonons drive transitions between the ground state $\ket{G1}$ and the its vibrational excited state $\ket{G1_{vib}}$ flipping its spin. From the fit, we extract $\Delta_{vib} \approx 7$~meV, consistent with the energies of the phonon-coupled states responsible for the PSB emission in the photoluminescence spectrum. Additionally, from the fitting parameters $C_R$ and $C_O$, we get values of $V^{(1)}_{vib}$ with a consistent order of magnitude of a few tens of meV.

Finally, we note that the data can also be fit by a power law model of the type $T_1^{-1} = \alpha T + \beta T^\gamma$, with $\gamma \approx 13$. Spin-lattice relaxation of this type has been previously reported for heavy ions in solid state environments \cite{kiel1967}. In that work, a Raman process is observed with a power dependence with $\gamma \approx 11 > 9$. They justify the large power observed by noting that the spin sublevels in the KD are not exactly each other's time-reversal conjugate, in such a way that the 'Van Vleck' cancellation does not happen completely \cite{abragam1970}. We exclude this as a relevant model in our case due to the fact that the power dependence necessary to explain our data is much higher, with $\gamma \approx 13$. Additionally, the consistency of $\Delta_{vib}$ observed by fitting the data with the energies observed in the PSB of the PL spectrum indicates that the rapid increase of the relaxation rate observed above $4$ K is indeed related to exponential Orbach processes involving $\ket{G1_{vib}\downarrow}$, $\ket{G1_{vib}\uparrow}$.